\documentclass[aps,twocolumn,showpacs]{revtex4}
\usepackage{amsmath}
\usepackage{epsfig}
\usepackage{booktabs} %调整表格线与上下内容的间隔
\usepackage{multirow}

\begin{document}

\title{Investigate the pentaquark resonance in the $NK$ system}
\newcommand*{\NJNU}{Department of Physics and Jiangsu Key Laboratory for Numerical
Simulation of Large Scale Complex Systems, Nanjing Normal University, Nanjing 210023, P. R. China}

\author{Xuejie Liu}\email{1830592517@qq.com}\affiliation{\NJNU}
\author{Hongxia Huang}\email{hxhuang@njnu.edu.cn}\affiliation{\NJNU}
\author{Jialun Ping}\email{jlping@njnu.edu.cn}\affiliation{\NJNU}

\begin{abstract}
A dynamical calculation of pentaquark systems with quark contents $uudd\bar{s}$ is performed in the framework
of quark delocalization color screening model with the help of resonating group method. The effective potentials
between baryon and meson clusters are given, and the possible bound states or resonances are investigated.
The single calculations show that the $NK^{*}$ with $I=0, J^{P}=\frac{1}{2}^{-}$, $\Delta K^{*}$ with
$I=1, J^{P}=\frac{1}{2}^{-}$, and $\Delta K^{*}$ with $I=2, J^{P}=\frac{3}{2}^{-}$ are all bound, but they all
turns into scattering states by coupling with the corresponding open channels. A possible resonance state
$\Delta K^{*}$ with $I=1, J^{P}=\frac{5}{2}^{-}$ is proposed. The mass is around $2110.5$ MeV, and the
decay modes are $NK$ in $D$-wave or $NK\pi\pi$ in $P$-waves.
\end{abstract}


\maketitle

\setcounter{totalnumber}{5}

\section{\label{sec:introduction}Introduction}
After decades of experimental and theoretical studies of hadrons, A lot of multiquark candidates have been proposed
for the hadrons beyond the ordinary quark-antiquark and three-quark structures. On one hand, the underlying theory
of the strong interaction, quantum chromodynamics (QCD) does not forbid the existence of the exotic hadronic states
such as glueballs (without quark/antiquark), hybrids (gluon mixed with quark/antiquark), compact multiquark states
and hadron molecules. On the other hand, dozens of nontraditional charmonium- and bottomonium-like states,
the so-called $XYZ$ mesons, have been observed during the past decades by the experimental
collaborations~\cite{x(3872),x(3872)1,x(3872)2,y(2175),y(2175)1,y(2175)2,y(2175)3,y(4220),y(4220)1,y(4220)2,y(4220)3,z(3900),z(3900)1}.

The intriguing pentaquark states were also searched in various colliders. In 2003, the LEPS collaboration announced
the observation of pentaquark $\Theta^{+}(1540)$~\cite{cta}, an exotic $K^{+}n$ or $K^{0}p$ resonance, which inspired
many theoretical and experimental work to search for pentaquarks. However, the existence of $\Theta^{+}(1540)$ is
not confirmed by other experimental collaborations~\cite{CLAS} and it is still a controversial issue~\cite{Nakano}.
Relatively, a study on pentaquarks was scarce to some extent until the observation of two candidates of hidden-charm
pentaquarks, $P^{+}_{c}(4380)$ and $P^{+}_{c}(4450)$ in the decay $\Lambda_{b}^{0}\rightarrow J/\psi K^{-}p$ by the
LHCb Collaborations~\cite{lhcb,lhcba,lhcbb}. A lot of theoretical calculations have been performed to investigate
these two exotic states~\cite{pc0,pc1,pc2,pc3,pc4,pc5,pc6,pc7,pc8,pc9,pc10}. In 2017, CERN announced an exceptional
new discovery that was made by the LHCb, which unveiled five new states all at one time~\cite{omeigc}.
These five states were also interpreted as exotic baryons~\cite{omeigc1,omeigc2,omeigc3}.

Now that the hidden charm pentaquarks were observed in the charmed sector, possible pentaquarks should also be
considered in the hidden strange sector, in which the $c\bar{c}$ is replaced by $s\bar{s}$. In fact, the $N\phi$
bound state was proposed by Gao {\em et~al.} in 2001~\cite{gao}. In Ref.~\cite{gao1}, the $N\phi$ resonance state
was obtained by channel coupling in the quark delocalization color screening model (QDCSM). Ref.~\cite{he.gao} showed
that a bound state could be produced from the $N\phi$ interaction with spin-parity $\frac{3}{2}^{-}$ after introducing
a Van der Waals force between the nucleon and $\phi$ meson. In Ref.~\cite{pc_like} the authors also studied possible
strange molecular pentaquarks composed of $\Sigma$ (or $\Sigma^{*}$) and $K$ (or $K^{*}$), and the results showed that
the $\Sigma K$, $\Sigma K^{*}$ and $\Sigma^{*} K^{*}$ with $IJ^{P}=\frac{1}{2}\frac{1}{2}^{-}$ and $\Sigma K^{*}$,
$\Sigma^{*} K$ and $\Sigma^{*} K^{*}$ with $IJ^{P}=\frac{1}{2}\frac{3}{2}^{-}$ were resonance states by coupling
the open channels. Besides, J. He interpreted the $N^{*}(1875)$ as a hadronic molecular states from the
$\Sigma^{*} K$ interaction~\cite{he.N}.

In addition to the hidden strange pentaquark, many theorists have also studied other possible pentaquark according to
the information of the experiment. For instance, the $\Lambda_{c}(2940)$ was reported by the BaBar Collaboration
by analyzing the $D^{0}p$ invariant mass spectrum~\cite{lan2940}, and it was confirmed as resonant structure in the
final state of $\Sigma_{c}(2455) \pi\rightarrow \Lambda_{c}\pi\pi$ by Belle~\cite{lan2940.1}. Since the
$\Lambda_{c}(2940)$ are near the threshold of $ND$, many works treat them as candidates of molecular states.
So there are a lot of work on $ND$ system. For example, Lifang {\em et~al.} did a bound state calculation of
$ND$ system in QDCSM and interpreted $\Lambda_{c}(2940)$ as a $ND^{*}$ molecular state~\cite{ND1}.
He {\em et~al.} also proposed that $\Lambda_{c}(2940)$ may be a $D^{*} p$ molecular state with
$J^{P}=\frac{1}{2}^{-}$~\cite{he1.ND}. Extending the study to the strange sector, we can also study the $NK$ system,
where the $D$ meson is replaced by the $K$ meson. In fact, many theoretical study have been devoted to the the $NK$
system. In Ref.~\cite{hiyama}, the authors use the standard non-relativistic quark model of Isgur-Karl to investigate
the $NK$ scattering problem, and the $NK$ scattering phase shift showed no resonance was seen in the energy region
$0-500$ MeV above the $NK$ threshold. In Ref.~\cite{NK}, Barns and Swanson used the quark-Born-diagram (QBD) method
to derive the $NK$ scattering amplitudes and obtained reasonable results for the $NK$ phase shifts, but they were
limited to $S-$wave. In Ref.~\cite{NK1}, the $NK$ interaction was studied in the constituent quark model and the
numerical results of different partial waves were in good agreement with the experimental date. Hence, it is
worthwhile to make a systematical study of $NK$ system by using different methods, which will deepen our
understanding about the possible pentaquarks.

It is a general consensus that it is difficult to directly study complicated systems in the low-energy region
by QCD because of the non-perturbative nature of QCD. So one has to rely on effective theories or QCD-inspired
models to tackle the problem of the multiquark. One of the common approaches to study the multiquark system
is the quark model. There are various kinds of the quark models, such as one-boson-exchange model, the chiral
quark model, the QDSCM, and so on. Particularly, the QDCSM was developed in the 1990s with the aim of explaining
the similarities between nuclear (hadronic clusters of quarks) and molecular force~\cite{QDCSM0,QDCSM1,QDCSM2}.
In this model, quarks confined in one cluster are allowed to escape to another cluster, this means that quark
distribution in two clusters is not fixed, which is determined by the dynamics of the interacting quark system,
thus it allows the quark system to choose the most favorable configuration through its own dynamics in a larger
Hilbert space. The confinement interaction between quarks in different clusters
is modified to include a color screening factor. The latter is a model description of the hidden-color channel-coupling
effect~\cite{QDCSM3}. This model has successful in describing nucleon-nucleon and hyperon-nucleon interactions and
the properties of the deuteron~\cite{sczb_s,QDCSM5,QDCSM4}. It is also employed to study the pentaquark system
in hidden-strange, hidden-charm, and hidden-bottom sectors~\cite{pc_like,pc_huang}.
%It is also employed to calculated the multiquark states scattering phase shifts and the candidates~\cite{qdcsm_pc,qdcsm_pc.shifs}. Recently, this model has been used to study the $P_{c}$-like pentaquarks~\cite{pc_like}. It was pointed out that the $\Sigma K$, $\Sigma K^{*}$, and $\Sigma^{*} K^{*}$ with $IJ^{P}$=$\frac{1}{2}\frac{1}{2}^{-}$ and $\Sigma K^{*}$, $\Sigma^{*} K$ and $\Sigma^{*} K^{*}$ with $IJ^{P}$=$\frac{1}{2}\frac{3}{2}^{-}$ are all resonance state by coupling the open channels, moreover, the molecular pentaquark $\Sigma^{*} K$ with quantum numbers $IJ^{P}$=$\frac{1}{2}\frac{3}{2}^{-}$ can be seen as a strange partner of the LHCb $P_{c}(4380)$ state.
In the present work, QDCSM is employed to study the nature of $NK$ systems, and the channel-coupling effect is
considered. Besides, we also investigate the scattering processes of the $NK$ systems to see if any bound or
resonance state exists or not.

This paper is organized as follows. In the next section, the framework of the QDCSM is briefly introduced. The
results for the $NK$ systems are shown in Sec. III, where some
discussion is presented as well. Finally, the summary is given in Sec. IV.

\section{THE QUARK DELOCALIZATION COLOR SCREENING MODEL (QDCSM)}
The quark delocalization, color screening model (QDCSM) is an extension of the native quark cluster model~\cite{naive}
and was developed with aim of addressing mutiquark systems.
The detail of QDCSM can be found in refs.~\cite{QDCSM0,QDCSM1,QDCSM2,QDCSM3,QDCSM5,QDCSM4}. Here, we just present
the salient features of the model. The model Hamiltonian is

\begin{widetext}
\begin{eqnarray}
H &=& \sum_{i=1}^{5} \left(m_i+\frac{\boldsymbol{p}_i^2}{2m_i}\right)-T_{CM}+\sum_{j>i=1}^5\left[V^C(r_{ij})+V^G(r_{ij})+V^B(r_{ij})\right],\\
%\iffalse
V^{G}(r_{ij}) &=& \frac{1}{4}\alpha_s \boldsymbol{\lambda}^{c}_i \cdot\boldsymbol{\lambda}^{c}_j
\left[\frac{1}{r_{ij}}-\frac{\pi}{2}\delta(\boldsymbol{r}_{ij})(\frac{1}{m^2_i}+\frac{1}{m^2_j}
+\frac{4\boldsymbol{\sigma}_i\cdot\boldsymbol{\sigma}_j}{3m_im_j})-\frac{3}{4m_im_jr^3_{ij}}
S_{ij}\right] \label{sala-vG} \\
V^{B}(r_{ij}) & = & V_{\pi}( \boldsymbol{r}_{ij})\sum_{a=1}^3\lambda_{i}^{a}\cdot \lambda
_{j}^{a}+V_{K}(\boldsymbol{r}_{ij})\sum_{a=4}^7\lambda_{i}^{a}\cdot \lambda _{j}^{a}
+V_{\eta}(\boldsymbol{r}_{ij})\left[\left(\lambda _{i}^{8}\cdot
\lambda _{j}^{8}\right)\cos\theta_P-(\lambda _{i}^{0}\cdot
\lambda_{j}^{0}) \sin\theta_P\right] \label{sala-Vchi1} \\
V_{\chi}(\boldsymbol{r}_{ij}) & = & {\frac{g_{ch}^{2}}{{4\pi}}}{\frac{m_{\chi}^{2}}{{\
12m_{i}m_{j}}}}{\frac{\Lambda _{\chi}^{2}}{{\Lambda _{\chi}^{2}-m_{\chi}^{2}}}}
m_{\chi} \left\{(\boldsymbol{\sigma}_{i}\cdot\boldsymbol{\sigma}_{j})
\left[ Y(m_{\chi}\,r_{ij})-{\frac{\Lambda_{\chi}^{3}}{m_{\chi}^{3}}}
Y(\Lambda _{\chi}\,r_{ij})\right] \right.\nonumber \\
&& \left. +\left[H(m_{\chi}r_{ij})-\frac{\Lambda_{\chi}^3}{m_{\chi}^3}
H(\Lambda_{\chi} r_{ij})\right] S_{ij} \right\}, ~~~~~~\chi=\pi, K, \eta, \\
V^C(r_{ij}) &=&  -a_{c}\boldsymbol{\lambda_{i}\cdot\lambda_{j}}[f(r_{ij})+V_{0}],\\
f(r_{ij}) & = &  \left\{ \begin{array}{ll}r_{ij}^2 &\qquad \mbox{if }i,j\mbox{ occur in the same baryon orbit} \\
\frac{1 - e^{-\mu_{ij} r_{ij}^2} }{\mu_{ij}} & \qquad \mbox{if }i,j\mbox{ occur in different baryon orbits} \\
\end{array} \right.\nonumber \\
S_{ij} &=& \left\{ \frac{(\boldsymbol{\sigma}_i
\cdot\boldsymbol{r}_{ij}) (\boldsymbol{\sigma}_j\cdot
\boldsymbol{r}_{ij})}{r_{ij}^2}-\frac{1}{3}\boldsymbol{\sigma}_i \cdot
\boldsymbol{\sigma}_j\right\},\\
H(x) &=& (1+3/x+3/x^{2})Y(x),~~~~~~
Y(x)=e^{-x}/x. \label{sala-vchi2}
%\fi
\end{eqnarray}
\end{widetext}
where $T_{cm}$ is the kinetic energy of the center-of-mass motion, and $\boldsymbol{\sigma}, \boldsymbol{\lambda}^c,
\boldsymbol{\lambda}^a$ are the SU(2) Pauli, SU(3) color, SU(3) flavor Gell-Mann matrices, respectively. $S_{ij}$ is
the quark tensor operator; The subscripts $i$, $j$ denote the
quark index in the system. The $Y(x)$ and $H(x)$ are the standard Yukawa functions~\cite{chiral}, the $\Lambda_{\chi}$ is the
chiral symmetry breaking scale, and the $\alpha_{s}$ is the effective scale-dependent running quark-gluon coupling
constant~\cite{oge}~,
$\frac{g_{ch}^2}{4\pi}$ is the chiral coupling constant for scalar and pseudoscalar chiral field coupling , determined from ~$\pi$~-nucleon-nucleon coupling constant through
\begin{equation}
\frac{g_{ch}^{2}}{4\pi}=\left(\frac{3}{5}\right)^{2} \frac{g_{\pi NN}^{2}}{4\pi} {\frac{m_{u,d}^{2}}{m_{N}^{2}}}
\end{equation}
In the phenomenological confinement potential $V^C$, the color screening parameter $\mu_{ij}$ is determined by fitting
the deuteron properties, $NN$ scattering phase shifts, and $N\Lambda$ and $N\Sigma$ scattering cross sections, respectively,
with $\mu_{qq}=0.45, \mu_{qq}=0.19$ and $\mu_{ss}=0.08$, satisfying the relation $\mu_{qs}^2= \mu_{qq}\mu_{ss}$
where $q$ represents $u$ or $d$.

The quark delocalization effect is realized by specifying the single-particle orbital wave function in QDCSM as
a linear combination of left and right Gaussians, the single-particle orbital wave functions used in the ordinary
quark cluster model are
\begin{eqnarray}
\psi_{\alpha}(\boldsymbol{s}_{i},\epsilon)&=&(\phi_{\alpha}(\boldsymbol{s}_{i})+\epsilon\phi_{\alpha}(\boldsymbol{-s}_{i}))/N(\epsilon),\\
\psi_{\beta}(\boldsymbol{s}_{i},\epsilon)&=&(\phi_{\beta}(\boldsymbol{-s}_{i})+\epsilon\phi_{\beta}(\boldsymbol{s}_{i}))/N(\epsilon),\\
N(\epsilon)&=& \sqrt{1+\epsilon^2+2\epsilon e^{-s^2_{i}/{4b^2}}},\\
\phi_{\alpha}(\boldsymbol{s}_{i})&=&(\frac{1}{\pi b^2})^{\frac{3}{4}}e^{-\frac{1}{2b^2}(\boldsymbol{r_{\alpha}}-\frac{2}{5}s_{i})^2},\\
\phi_{\beta}(\boldsymbol{-s}_{i})&=&(\frac{1}{\pi b^2})^{\frac{3}{4}}e^{-\frac{1}{2b^2}(\boldsymbol{r_{\beta}}+\frac{3}{5}s_{i})^2},
\end{eqnarray}
The $\boldsymbol{s}_{i}$, $i=1,2,..., n$, are the generating coordinates, which are introduced to expand the
relative motion wave function~\cite{QDCSM1,QDCSM2,sczb_s}. The mixing parameter $\epsilon(s_{i})$ is not an adjusted
one but determined variationally by the dynamics of the multi-quark system itself. It is this assumption that allows
the multi-quark system to choose its favorable configuration in the interacting process. It has been used to explain the
cross-over transition between the hadron phase and the quark-gluon plasma phase~\cite{phase}. All the other symbols
in the above expressions have their usual meanings. All the parameters of the Hamiltonian are from our previous work
of hidden strange pentaquark~\cite{pc_like}.

\section{The results and discussions}
In this work, we investigate the $NK$ systems with $I=0, 1, 2, J^{P}=\frac{1}{2}^{-}, \frac{3}{2}^{-},
\frac{5}{2}^{-}$ in the QDCSM. For the negative-parity, the orbital angular momentum $L$ between two clusters is
set to $0$. All the channels involved are listed in Table~\ref{channels}. To investigate the properties of the
$NK$ systems and to see if any bound or resonance state exists or not, three steps are invoked.

\begin{table}[ht]
\caption{\label{kkk}The coupling channels of each quantum number.}
\begin{tabular}{ccccc}
 \hline \hline
% \multicolumn{5}{r}{channels} \\
 ~~~$I=0$ &$s=\frac{1}{2}$~~~  &\multicolumn{3}{c}{$NK$,~~$NK^{*}$}\\
 ~~~$I=1$ &$s=\frac{1}{2}$~~~  &\multicolumn{3}{c}{$NK$,~~$NK^{*}$,~~$\Delta K^{*}$}\\
 ~~~$I=1$ &$s=\frac{3}{2}$~~~  &\multicolumn{3}{c}{$NK^{*}$,~~$\Delta K$,~~$\Delta K^{*}$}\\
 ~~~$I=1$ &$s=\frac{5}{2}$~~~  &\multicolumn{3}{c}{$\Delta K^{*}$}\\
 ~~~$I=2$ &$s=\frac{1}{2}$~~~  &\multicolumn{3}{c}{$\Delta K^{*}$}\\
 ~~~$I=2$ &$s=\frac{3}{2}$~~~  &\multicolumn{3}{c}{$\Delta K$,~~$\Delta K^{*}$}\\
 ~~~$I=2$ &$s=\frac{5}{2}$~~~  &\multicolumn{3}{c}{$\Delta K^{*}$}\\

 \hline\hline
\end{tabular}
\label{channels}
\end{table}

\subsection{The effective potential calculation}

\begin{figure}[b]
\epsfxsize=3.7in \epsfbox{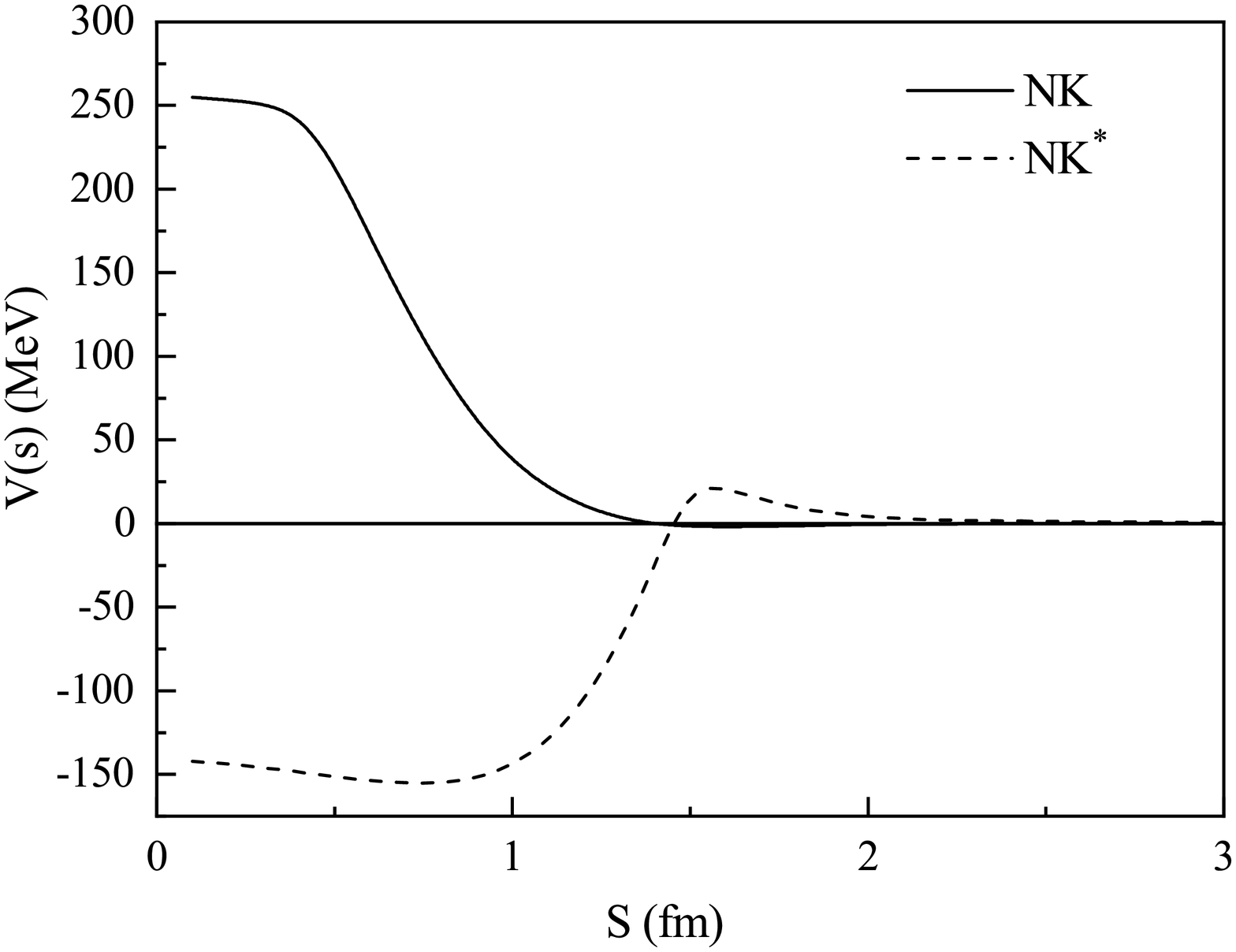}
\caption{The effective potential of different channels for the $NK$ system with $I=0$.}
\end{figure}

Because the attractive potential is necessary for forming a bound state or a resonance, for the first step, the effective
potentials of all the channels listed in the Table~\ref{channels} are calculated. The effective potential between two
colorless clusters is defined as,
$$
V(s)=E(s)-E(\infty),
$$
where $E(s)$ is the energy of the state at the separation
$s$ between two clusters. The effective potentials of the $S$-wave $NK$ systems with $I=0, 1, 2$ are shown in
Figs. 1-3, respectively. For the $IJ^{P}=0\frac{1}{2}^{-}$ system (Fig. 1), one see that the potential
of the $NK$ state is almost repulsive, which means that the $NK$ is difficult to form a bound state, while the potential
of the $NK^{*}$ channel is attractive in the short range, a bound state or a resonance $NK^{*}$ is possible.
For the $I=1$ system, Fig. 2(a) shows the potential of the $NK$ system with $J^{P}=\frac{1}{2}^{-}$, in which the
potential of the channel $NK$ shows repulsive property, while other two channels are attractive.
The attraction between $\Delta$ and $K^{*}$ is much larger than that of the $NK^{*}$ channel,
which indicates that it is possible for $\Delta K^{*}$ to form a bound or resonance state. In Fig. 2(b), the potentials
of both the $J^{P}=\frac{3}{2}^{-}$ channel $\Delta K$ and $\Delta K^{*}$ are weakly attractive and the potential of
the channel $NK^{*}$ is repulsive. From Fig. 2(c), it is obvious that the potential of the $J^{P}=\frac{5}{2}^{-}$ channel
$\Delta K^{*}$ has a strong attraction, it is interesting to explore the possibility of formation of bound or resonance state.
For the $I=2$ system, the potential of both the $J^{P}=\frac{1}{2}^{-}$ and
$\frac{3}{2}^{-}$ $\Delta K^*$ channels are attractive, a dynamic calculation is needed here to check the existence
of bound or resonance states. The potentials of the $\Delta K$ with the
$J^{P}=\frac{3}{2}^{-}$ and the $\Delta K^{*}$ with the $J^{P}=\frac{5}{2}^{-}$ are repulsive, bound or resonance state
is impossible here.

\begin{figure}[t]
\epsfxsize=3.5in \epsfbox{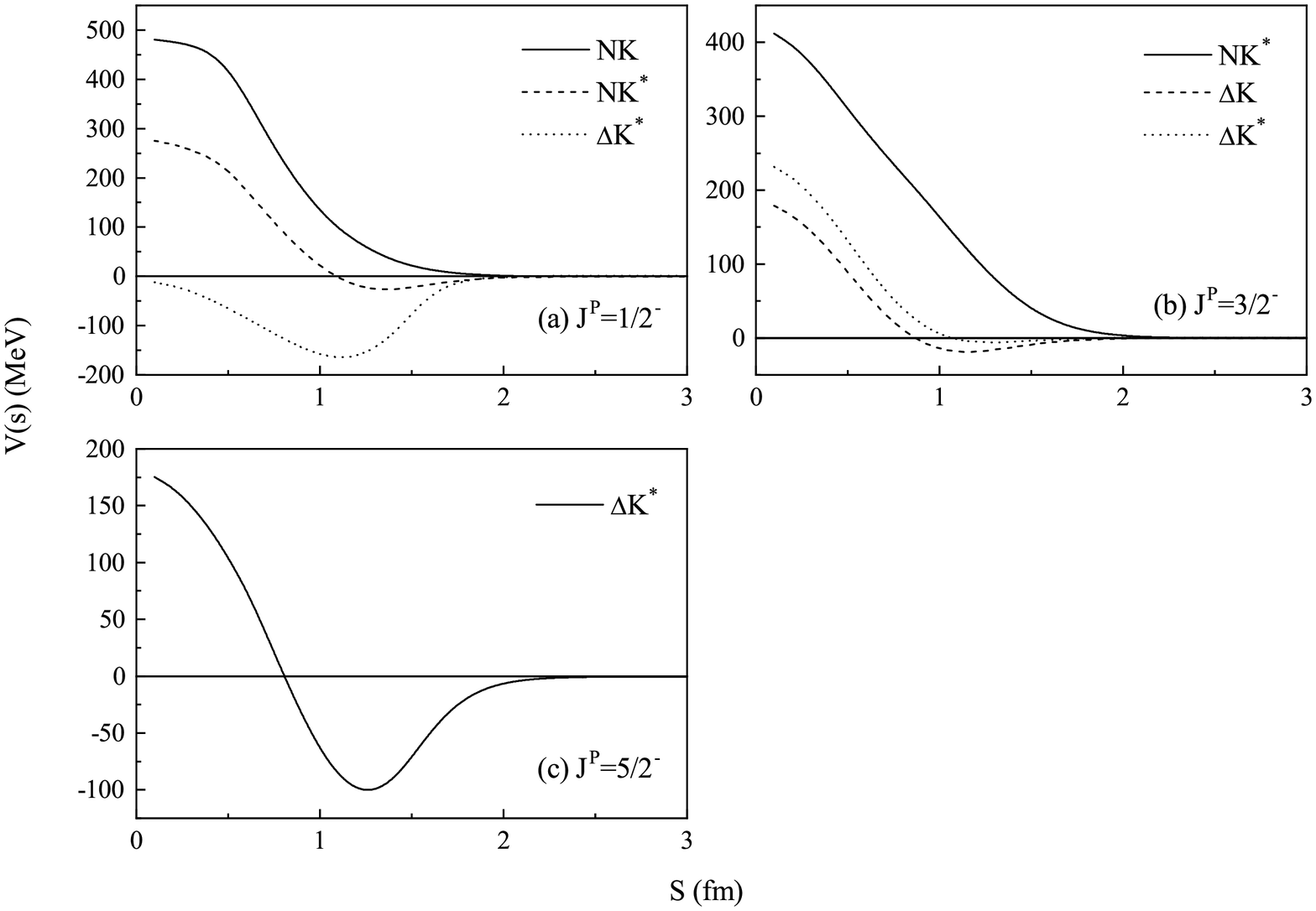}
\caption{The effective potential of different channels for the $NK$ system with $I=1$.}

\epsfxsize=3.5in \epsfbox{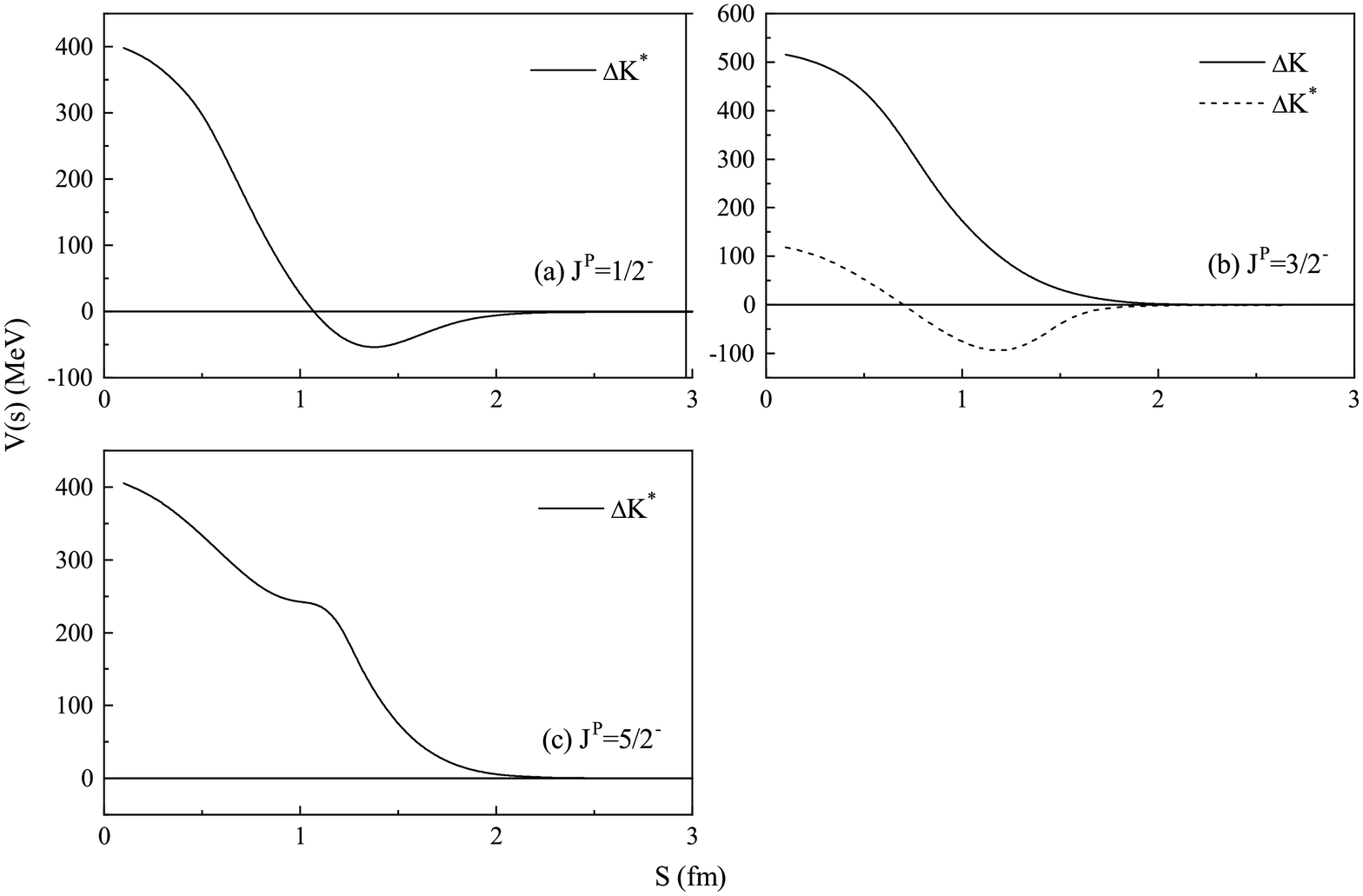}
\caption{The effective potential of different channels for the $NK$ system with $I=2$.}
\end{figure}

\subsection{The bound state calculation}

In order to check whether the possible bound or resonance states can be realized, a dynamic calculation is needed.
Here the RGM equation, which is a successful method in nuclear physics for studying a bound-state problem or
scattering one, is employed. Expanding the relative motion wave function between two clusters by Gaussians,
then the integro-differential equation of the RGM can be reduced to a algebraic equation, the generalized
eigen-equation. The energy of the system can be obtained by solving the eigen-equation. The details of solving
the RGM equation can be found in Ref.~\cite{RGM0,RGM1}. In the calculation, the baryon-meson separation is taken
to be less than 6 fm (to keep the matrix dimension manageably small). The binding energies and the masses of
every single channel and those with channel coupling are listed in Table~\ref{bound1}.
\begin{table*}[ht]
\caption{The binding energies and the masses of every single channels and those of channel coupling for
the molecular pentaquarks. The values are provided in units of MeV, $ub$ and $-$ represent unbound and
the channel does not exist, respectively}
\begin{tabular}{lccccccc}
\hline \hline
Channel &
$IJ^{P}$=$0\frac{1}{2}^{-}$ &$IJ^{P}$=$1\frac{1}{2}^{-}$  &$IJ^{P}$=$1\frac{3}{2}^{-}$ &$IJ^{P}$=$1\frac{5}{2}^{-}$ &$IJ^{P}$=$2\frac{1}{2}^{-}$ &$IJ^{P}$=$2\frac{3}{2}^{-}$ &$IJ^{P}$=$2\frac{5}{2}^{-}$ \\
$NK$ & ub & ub &-  &-  &-  &-  &-  \\
$NK*$ & -62.3$/1768.7$ & ub & ub  &-  &-  &-  &-  \\
$\Delta K$  & - & - &ub  &-  &-  &ub  &-  \\
$\Delta K*$  & - & -68.1/2055.9 &ub  &-13.5/2110.5 &ub &-10.2/2113.8 &ub  \\
$E_{cc}$& ub & ub & ub & bound  & ub  & ub  & ub   \\

\hline
\end{tabular}
\label{bound1}
\end{table*}

For the $I=0, J^{P}=\frac{1}{2}^{-}$ system, the single channel calculation shows that the energy of the $NK$ channel
is above the threshold because the attraction between $N$ and $K$ is too weak to tie the two particles together,
which means that there is no bound state in this channel. However, for the $NK^{*}$ state, the strong attractive
interaction between $N$ and $K^{*}$ leads to the energy of the $NK^{*}$ state below the threshold of the two particles,
so the $NK^{*}$ state is bound in the single channel calculation. By coupling two channels of $NK$ and $NK^{*}$,
the lowest energy is still above the threshold of the $NK$ channel, which indicates that no bound state for
$I=0, J^{P}=\frac{1}{2}^{-}$ system. However, we should check if the $NK^{*}$ is a resonance state in the
channel coupling calculation, which is presented in the next sub-section.

For the $I=1$ system, the state with $J^{P}=\frac{1}{2}^{-}$ has three channels: $NK$, $NK^{*}$, and $\Delta K^{*}$.
The $NK$ and $NK^{*}$ are all unbound. It is reasonable. As shown in Fig.2(a), the effective potential between
$N$ and $K$ is repulsive, and the one between $N$ and $K^{*}$ is weakly attractive. So neither $NK$ nor $NK^{*}$
is bound here. However, the attraction between $\Delta$ and $K^{*}$ is strong enough to bind $\Delta$ and $K^{*}$,
so the $\Delta K^{*}$ is a bound state with the binding energy of $-68.1$ MeV in the single calculation.
Then the channel-coupling is also considered. The lowest energy is still is above the threshold of the $NK$ channel
and it means that there is no bound state for $I=1$ $J^{P}=\frac{1}{2}^{-}$ system.
The $\Delta K^{*}$ may turn out to be a resonance state by coupling to the open channels, $NK$ and $NK^{*}$,
which should be investigated in the scattering process of the open channels.
The state with $J^{P}=\frac{3}{2}^{-}$ includes three channels: $NK^{*}$, $\Delta K$and $\Delta K^{*}$.
The effective potential of $NK^{*}$ is repulsive which make the state unbound. Both the $\Delta K$ and $\Delta K^{*}$ are
also unbound due to the weakly attractive potentials between $\Delta$ and $K$ or $K^{*}$ as shown in Fig.2(b).
The coupling of all channels also cannot make any state bound.
For the $J^{P}=\frac{5}{2}^{-}$ system, there is only one channel: $\Delta K^{*}$. The attraction between
$\Delta$ and $K^{*}$ is large enough to form a bound state, and the binding energy is $-13.5$ MeV.

For the $I=2$ system, both $\Delta K^{*}$ with $J^{P}=\frac{1}{2}^{-}$ and $J^{P}=\frac{5}{2}^{-}$ are unbound.
For the $J^{P}=\frac{3}{2}^{-}$ system, the $\Delta K$ is unbound while the $\Delta K^{*}$ is bound with the binding
energy of $-10.2$ MeV in the single channel calculation. However, the channel-coupling cannot push the lowest energy
under the threshold of the $\Delta K$ channel. So no bound state is obtained by channel-coupling. We will check if
$\Delta K^{*}$ is a resonance state by coupling the open channel.

It is worth to mention that a subtraction procedure is used here to obtain the mass of a bound state here. Because
the quark model cannot reproduce the experimental masses of all baryons and mesons, the theoretical threshold and
the experimental threshold for a given channel is different (the threshold is the sum of the masses of the baryon
and the meson in the given channel). However, the binding energy, the difference between the calculated energy of
the state and the theoretical threshold can minimize the deviation. So we define the mass of a bound state as
$M=M^{cal}(5q)-M^{cal}(B)-M^{cal}(M)+M^{exp}(B)+M^{exp}(M)$, where $M(B)$ and $M(M)$ denote the baryon mass and the
meson mass, respectively, and the superscripts $cal$, $exp$ stand for the calculated and experimental.

\subsection{The resonance state calculation}
Resonances are unstable particles usually observed in the scattering process. The bound state in the single channel
calculation may turn to be a resonance after coupling with open channels. Here, we calculate the baryon-meson scatting
phase shifts and investigate the resonance states by using the RGM.

From the bound state calculation showed above, for the $I=0, J^{P}=\frac{1}{2}^{-}$ system, the single channel $NK^{*}$
is bound, while the $NK$ channel is unbound and is identified as the open channel. For the $I=1, J^{P}=\frac{1}{2}^{-}$
system, there are two open channels ($NK$,$NK^{*}$) and one bounded channel ($\Delta K^{*}$).
For the $I=2, J^{P}=\frac{3}{2}^{-}$ system, it is similar to the $I=0, J^{P}=\frac{1}{2}^{-}$ system.
The open channel and the bounded channel is $\Delta K$ and $\Delta K^{*}$, respectively. Here, we only consider the
channel-coupling in $S-$wave, which is through the central force. The channel-coupling between the $S-$ and $D-$ wave states
is very small, which is through the tensor force, and is ignored here. All the scattering phase shifts of the open channels
are shown in Fig. 4.

For the $I=0, J^{P}=\frac{1}{2}^{-}$ system, there is no any resonance state appeared in the phase shifts of
the open channel $NK$, which means that the bound state $NK^{*}$ in the single channel calculation turns into scattering
state after coupling with the $NK$ channel. The case is similar for both the $I=1, J^{P}=\frac{1}{2}^{-}$ system and
the $I=2, J^{P}=\frac{3}{2}^{-}$ system. As shown in Fig .4(b), no resonance state appeared in the phase shifts of the
open channel $NK$ or $NK^{*}$, which indicates that the bound state $\Delta K^{*}$ with $I=1, J^{P}=\frac{1}{2}^{-}$ is not
a resonance state by coupling with the open channels. In Fig .4(c), we can also see that $\Delta K^{*}$ with
$I=2, J^{P}=\frac{3}{2}^{-}$ is not a resonance by coupling to the open channel $\Delta K$.

\begin{figure}
\epsfxsize=3.5in \epsfbox{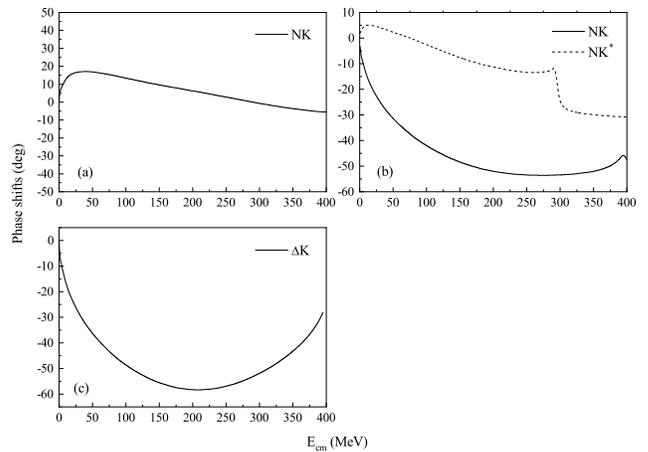} \vspace{-0.2in}

\caption{The phase shift of the (a) $I=0, J^{P}=\frac{1}{2}^{-}$, (b) $I=1, J^{P}=\frac{1}{2}^{-}$, (c) $I=2, J^{P}
=\frac{3}{2}^{-}$.}
\end{figure}

\section{Summary}
In the framework of the QDCSM, the pentaquark systems with quark contents $uudd\bar{s}$ are investigated by means of
RGM. All the effective potentials between baryon and meson are calculated to search for the strong attraction, which is
the necessary condition for forming bound state or resonance. The dynamic calculation show that the states $NK^{*}$ with
$I=0, J^{P}=\frac{1}{2}^{-}$, $\Delta K^{*}$ with $I=1, J^{P}=\frac{1}{2}^{-}$, and $\Delta K^{*}$ with
$I=2, J^{P}=\frac{3}{2}^{-}$ are all bound in the single channel calculation due to the strong attraction of the states.
However, all these bound states turns into scattering states by coupling with the open channels. It indicates that
the effect of the coupling with the open channels cannot be neglected, because it will transfer the bound state into
a resonance state or a scattering state. There is only one bound state in our calculation, which is the $\Delta K^{*}$
with $I=1, J^{P}=\frac{5}{2}^{-}$ with the energy of $2110.5$ MeV. However, in present calculation, we only consider all
possible channels in $S-$wave. The $D$-wave $\Delta K$ channel can couple to $\Delta K^{*}$ through the tensor interaction.
The coupling is expected to turn the bound state to a resonance with decay width of several MeV, which is our next work. 
The $\Delta K^{*}$ state can
also decay to $NK\pi\pi$ in $P$-waves (two $P$-waves are needed to conserve the parity).

\acknowledgments{}
This work is supported partly by the National Science Foundation
of China under Contract Nos. 11675080, 11775118 and 11535005, the Natural Science Foundation of
the Jiangsu Higher Education Institutions of China (Grant No. 16KJB140006).

\end{document}